\documentclass[prb,aps,twocolumn,preprintnumbers,superscriptaddress,longbibliography]{revtex4-1}

\usepackage{latexsym,epsfig,amssymb,amsfonts,amsmath,graphicx,color,bm,times,hyperref,amstext,epstopdf}

\usepackage{amsmath,amssymb,amsfonts,amstext}
\usepackage{graphicx,color,sidecap}          
\usepackage{epsfig}                  
\usepackage{wrapfig}
\usepackage{subfigure}
\usepackage{longtable}               
\usepackage{makeidx}                 
\usepackage{mathrsfs}
\usepackage{amsfonts}
\usepackage[T1]{fontenc}		
\usepackage[utf8]{inputenc}		
\usepackage{indentfirst}		
\usepackage{color}				
\usepackage{graphicx}			
\usepackage{microtype} 			
\usepackage[Lenny]{fncychap}
\usepackage{fancyhdr}
\usepackage{url}
\usepackage{soul}

\begin{document}

\title{Energy current manipulation and reversal of rectification in graded XXZ spin
chains}

\author{Alberto L. de Paula Jr.}
\affiliation{Departamento de F\'isica, Universidade Federal de Minas Gerais,
C. P. 702, 30123-970, Belo Horizonte, MG, Brazil}
\affiliation{Instituto Federal de Educação, Ciência e Tecnologia do Rio de Janeiro,
C. P. 121, 20270-021, Rio de Janeiro, RJ, Brazil}
\author{Emmanuel Pereira}
\affiliation{Departamento de F\'isica, Universidade Federal de Minas Gerais,
C. P. 702, 30123-970, Belo Horizonte, MG, Brazil}
\author{Raphael C. Drumond}
\affiliation{Departamento de Matem\'atica, Universidade Federal de Minas Gerais,
C. P. 702, 30123-970, Belo Horizonte, MG, Brazil}
\author{M. C. O. Aguiar}
\affiliation{Departamento de F\'isica, Universidade Federal de Minas Gerais,
  C. P. 702, 30123-970, Belo Horizonte, MG, Brazil}


\begin{abstract}
This work is devoted to the investigation of nontrivial transport properties in many-body
quantum systems. Precisely, we study transport in the steady state of spin-1/2 Heisenberg XXZ
chains, driven out of equilibrium by two magnetic baths at their end points. We take graded
versions of the model, i.e., asymmetric chains in which some structure gradually changes in space.
We investigate how we can manipulate and control the energy and spin currents of such chains by
tuning external and/or inner parameters.  In particular, we describe the occurrence of energy
current rectification and its reversal due to the application of external magnetic fields. We
show that, after carefully chosen inner parameters for the system, by turning on an external
magnetic field we can find spin and energy currents propagating in different directions. More
interestingly, we may find cases in which rectifications of energy and spin currents
occur in opposite directions, i.e., if the energy current is larger when flowing from left to right
side, then the spin current is larger if it flows from right to
left side. We still describe situations with inversion of the energy current direction as we
increase the system asymmetry.
We stress that our work aims the development of theoretical knowledge as
well as the stimulation of future experimental applications.
\end{abstract}

\maketitle

\section{Introduction}

The exchange of matter and energy among systems and environments is the
portrait of non-equilibrium statistical physics: consequently, one of its keystone
is the understanding of the transport laws. Photonics, spintronics,
phononics, in addition to electronics are important research fields devoted
to the theoretical study, as well as to the experimental manipulation and control
of different forms of transport. In particular, achievements of electronics have
had great impact in our daily lives, due to the invention of transistor and other
related devices whose basic component is the diode. The electronic diode, or
rectifier, is a system in which the flow of electric charges has a preferential
direction. Inspired by such a success, a considerable effort has been dedicated
to the search of other feasible rectifiers, such as those based on the heat and
spin currents. In phononics, for example, in spite of the absence of an efficient
thermal diode, a list of familiar electronics analogs has been theoretically
proposed: thermal transistors, thermal logical gates, thermal memories, among
others.\cite{baowen_li_2012_rmp} It is interesting to note that most of
these proposals involve classical anharmonic chains of oscillators, a recurrent
model for the study of heat conduction in solids, used since Debye and Peierls.
Motivated by these works, a solid state thermal diode has been already experimentally
built, given by a carbon and boron nitride nanotube, inhomogeneously coated with
heavy molecules, see Ref.~\onlinecite{Chang_2006_science}. However, such a diode
presents a very small rectification factor.

In this context, but with focus on the transport properties in low-dimensional
quantum systems, the present paper addresses the investigation of graded quantum
XXZ open chains, boundary driven by magnetization baths. The present ambient of
miniaturization together with the scarceness of related results (e.g., about details
of the rectification properties) makes important the analysis of genuine quantum
models. Besides the importance of the detailed investigation of transport and
rectification in general quantum systems, it is worth to stress the relevance
and the recent interest in the study of the one dimension (1D) XXZ chain by
itself. It is the archetypal model to the analysis of open quantum
systems,\cite{petruccione} drawing attention to important problems in different
areas, such as optics, non-equilibrium statistical physics, condensed matter,
quantum information and cold-atoms. Interestingly, it has been experimentally
realized.\cite{Trotzky_2008_science,barreiro_2011_nature,hauke2010complete,bloch2012quantum,blatt2012quantum,toskovic2016atomic}
For instance, in Ref.~\onlinecite{Trotzky_2008_science} it is reported a direct
observation of superexchange interactions (the magnitude and sign of which can
be controlled) between atoms in different sides of an optical double well.
More recently, the XXZ chain in a transverse field was realized by manipulating
atoms using scanning tunneling microscopy.~\cite{toskovic2016atomic}

In the present work, we analyze boundary driven XXZ models, i.e., systems with
target polarization at the edges. We do not consider the case of chains
weakly and passively coupled to thermal baths, such as the spin systems considered
in Refs.~\onlinecite{baowen_li_2009_1,baowen_li_2009_2}. The existence of many
real situations, exemplified by information processes using feedback control and
reservoirs prepared in non-equilibrium, force us to go beyond the models of driven
systems weakly coupled to thermal baths. The boundary driven case, which is
the one we consider, can be phenomenologically justified in terms
of the ``repeated interactions'' protocol \cite{karevski_2009_prl} and other
schemes\cite{petruccione}, and has been used in many recent works of the
literature.\cite{emmanuel_2016_pre,emmanuel_2017_pre,landi_2014_pre,prosen2010exact,prosen2011exact,popkov_2013_pre,Medvedyeva_2016_prb,buvca2014exactly,prosen2011open,benenti2009negative}
A detailed study of the thermodynamics behind these systems associated to the
repeated interactions protocol is described in Ref.~\onlinecite{strasberg_2017_prx}.
It is interesting to note that dissipators such as the ones we consider can be
experimentally realized and controlled in cold atoms
setups.\cite{schwager2013dissipative, ramos2014quantum} Moreover, spin and thermal
reservoirs for quasi-one-dimensional clouds can be implemented and the subsequent
currents measured.\cite{krinner2016mapping, krinner2017two}


In the current work we are motivated, as previously said, by the present interest
in understanding and controlling transport properties of many-body quantum chains.
Specifically, we are stimulated by previous works \cite{emmanuel_2016_pre,emmanuel_2017_pre}
in which a ubiquitous occurrence of energy current rectification has been shown for
graded XXZ chains with or without the presence of an uniform external magnetic field.
Here, we investigate how we can manipulate and control the energy and spin flows by
tuning external and/or inner parameters, such as an external magnetic field or the
system asymmetry. In particular, we describe systems with specific
inner parameters for which current rectification is observed. Then, we show how to
manipulate the currents by the addition of properly chosen magnetic fields, with
the goal of reversing their rectification. We describe how the spin and energy
currents change as a function of parameters such as the inter-site interaction, the
system size, and the system degree of asymmetry. We aim, besides the achievement of
theoretical knowledge, to inspire possible experimental applications.

The paper is organized as follows. In Section II we detail the
  model describing both the bulk of the chain and the reservoirs, as well as the
  expressions for the spin and energy currents we are interested in. Next, we
  present our results: in subsection III.A we discuss the behavior of the
  currents as a function of the inter-site interaction; subsection III.B is
  devoted to the system behavior when its degree of asymmetry changes; and
  finally, in subsection III.C, we compare
  results for chains in different ``phases'' (we use quotation marks to emphasize
  that phases are rigorously defined only in the thermodynamic limit). In
section IV we summarize our conclusions.

\


%

\section{Model}\label{model}

In this work we study nontrivial properties of the spin and energy currents in a
spin-1/2 chain described by the Heisenberg XXZ model, which is connected to magnetic
reservoirs present at its edges. The Hamiltonian of the chain is given by

\begin{eqnarray}\label{XXZ}
  H&=&\sum_j^{N-1}\left[\alpha \left(\sigma_j^X\sigma_{j+1}^X + \sigma_j^Y\sigma_{j+1}^Y
    \right) \right. \nonumber \\
    &+& \left.  \Delta_{j,j+1}\sigma_j^Z\sigma_{j+1}^Z  \right] +
  \sum_j^N B_j\sigma_j^Z, \label{xxz}
 \end{eqnarray}
where $\sigma^{\beta}_j$ ($\beta=X,Y,Z$) are the Pauli matrices on site $j$; $\alpha$
is the exchange coupling constant; $\Delta_{j,j+1}$ is the anisotropy parameter;
$B_i$ is an external magnetic field acting on site $j$; and $N$ is the number of
sites in the chain. Throughout this paper we use $\alpha=1$ as the unity of energy.
Note that both $B_j$ and $\Delta_{j,j+1}$ can vary along the chain.



We consider a Markovian dynamics so the density matrix
$\rho$ of the chain can be described by the master equation in the Lindblad form \cite{petruccione}
\begin{equation}
 \frac{d\rho}{dt}=-i[H,\rho]+\mathcal{D}[\rho],
\end{equation}
where we set $\hbar=1$, and $\mathcal{D}[\rho]$ satisfies
\begin{equation}\label{diss}
		\mathcal{D}[\rho]=\sum_k \left(L_k\rho L_k^{\dagger}-\frac{1}{2}\{L_k^{\dagger}L_k,\rho\}\right) .
\end{equation}
Here,  $\{A,B\}$ denotes the anticommutator between $A$ and $B$, and
the Lindblad operators $L_k$ are given by
\begin{eqnarray}
  L_{1}&=&\sqrt{\frac{\varepsilon}{2}(1 + f)}\sigma^{+}_1, ~~~~ L_{2}=\sqrt{\frac{\varepsilon}{2}(1 - f)}\sigma^{-}_1,  \label{lind2_1} \\
  L_{3}&=&\sqrt{\frac{\varepsilon}{2}(1 - f)}\sigma^{+}_N,  ~~~~ L_{4}=\sqrt{\frac{\varepsilon}{2}(1 + f)}\sigma^{-}_N, \label{lind2_2}
\end{eqnarray}
where $\sigma^{\pm} = \frac{1}{2}\left(\sigma^{X} \pm i \sigma^{Y}\right)$
are raising and lowering operators and $\varepsilon$ measures
the coupling between the reservoirs and the chains.
In the equations above,
$L_k$ are written for the particular case in which $f$
(quantity that weights how much each
operator acts at the chain edges) has the same
magnitude but opposite signs for the
left and right reservoirs - for more details, see, for
example, Refs. \onlinecite{landi_2014_pre,emmanuel_2016_pre}.

Note that there are four Lindblad operators, two for each reservoir,
and that the operators are applied at the first ($i=1$) and the last
($i=N$) chain sites. To be more specific, in each chain edge there
exists an operator $\sigma^{+}$, with
weights given by $\sqrt{\frac{\varepsilon}{2}(1 + f)}$ and
$\sqrt{\frac{\varepsilon}{2}(1 - f)}$ for the first and the
last site, respectively, which tries to ``induce'' the spin to
have projection {\it up} in the $Z-$direction.
In each edge, it also exists an operator
$\sigma^{-}$, which tries to ``force'' the spin to point {\it down},
now with weights given by $\sqrt{\frac{\varepsilon}{2}(1 - f)}$ and
$\sqrt{\frac{\varepsilon}{2}(1 + f)}$ for the first and the
last site.

The Lindblad form is quite general, since a differential equation is
a Markovian master equation if and only if it can be written in the
Lindblad form.\cite{petruccione} moreover, as mentioned in the introduction,
the specific boundary-driven case treated here can be phenomenologically justified in terms of the ``repeated
interactions'' protocol.\cite{karevski_2009_prl}


\subsection{Asymptotic state and currents}

Once the reservoirs are connected to the chain, the problem involves an
out-of-equilibrium state, and a natural question is which is
the asymptotic state of the system. Typically these reservoirs do not the thermalize the chain and asymptotically it goes to a non-equilibrium steady state (NESS). \cite{prosen2010exact,prosen2011exact} In both numerical
and analytic recent works, there is an interest in studying
properties of the NESS, such as the magnetization and
transport quantities in spin systems attached to magnetic
reservoirs \cite{popkov_2013_pre,Medvedyeva_2016_prb,buvca2014exactly,prosen2011open,benenti2009negative,mendoza2018asymmetry,vznidarivc2011spin,vznidarivc2016diffusive,PhysRevLett.120.200603} or the
occupation profile when considering a chain described by the
Hubbard model, coupled to reservoirs that allow for the exchange
of particles. \cite{pizorn_2013_pra,prosen_2014_prl,bonnes_2014_pra}

The NESS is defined by imposing the condition
\begin{equation}
\frac{d \rho_{\infty}}{dt}=0,
\end{equation}
where $\rho_{\infty}=\rho(t \rightarrow \infty)$. For spin
chains coupled to magnetic reservoirs, which is the case we
analyze here, one can study the magnetization profile, given by
$\left< \sigma_j^Z \right> = tr\{\sigma_j^Z \rho_{\infty} \}$, as
well as the behavior of the spin and energy currents as a
function of $B_j$ and the anisotropy parameter $\Delta_{j,j+1}$.

The spin current is given by $J^S=\left< J_j^S \right> = tr\{J_j^S
\rho_{\infty} \}$, where
\begin{eqnarray}
  J_j^S&=&4 i\left( \sigma_j^{+}\sigma_{j+1}^{-} -
  \sigma_j^{-}\sigma_{j+1}^{+} \right) \nonumber \\
  &=&  2 \left( \sigma_j^X\sigma_{j+1}^Y - \sigma_j^Y\sigma_{j+1}^X
  \right)
\end{eqnarray}
is the current on site $j$ in the bulk of the chain ($2 \leq j \leq N-1$), obtained from the
continuity equation $d \left< \sigma_j \right>/dt
= \left< J_{j-1}^S \right> - \left< J_{j}^S \right>$.\cite{emmanuel_2016_pre}



If we define $E_{j,j+1}$ such that Eq.~(\ref{XXZ})
is written as $H = \sum_{j=1}^{N-1} E_{j,j+1}$, the energy
current $\left< J_{j}^E \right>$ can be defined from the
following continuity
equation\cite{mendoza-arenas_2013,emmanuel_2016_pre}
\begin{equation}
\frac{d \left< E_{j,j+1} \right>}{dt}
= \left< J_{j}^E \right> - \left< J_{j+1}^E \right>.
\end{equation}
The Hamiltonian we consider can be divided into two terms, the
XXZ one, which involves the coupling between the spins in
neighboring sites, and a local term, proportional to the
magnetic field $B_j$. From the continuity equation above, one can
show\cite{mendoza-arenas_2013,emmanuel_2016_pre} that the energy
current has two contributions, one coming from the XXZ term and
another coming from the presence of the external magnetic field,
which we label with superscript $B$,
 \begin{equation}\label{cor-energ}
  \langle J_j^E\rangle=\langle J_j^{XXZ} \rangle+\langle J_j^B \rangle,
 \end{equation}
where
  \begin{eqnarray}\label{XXZ-corrente}
    \langle J_j^{XXZ} \rangle &=& 2 \langle \left(\sigma_{j-1}^Y\sigma_j^Z\sigma_{j+1}^X
    - \sigma_{j-1}^X\sigma_j^Z\sigma_{j+1}^Y\right) \\ \nonumber
    &+&\Delta_{j-1,j}\left(\sigma_{j-1}^Z\sigma_j^X\sigma_{j+1}^Y
    - \sigma_{j-1}^Z\sigma_j^Y\sigma_{j+1}^X\right) \\ \nonumber
    &+&\Delta_{j,j+1}\left(\sigma_{j-1}^X\sigma_j^Y\sigma_{j+1}^Z
    - \sigma_{j-1}^Y\sigma_j^X\sigma_{j+1}^Z\right) \rangle \nonumber
 \end{eqnarray}
and
  \begin{equation}\label{campo}
  \langle J_j^B \rangle=\frac{B_j}{2}\langle J_{j-1}^S + J_j^S \rangle.
 \end{equation}

  \noindent
Note that, in the steady state which we are interested in, the currents
are uniform along the chain, therefore the spin and energy currents
do not depend on the pair of sites considered.

It is important to stress that if the system is {\it homogeneous}
(this is the case usually considered in the literature, but not in
the present paper), then $\langle J_i^{XXZ} \rangle=0$.\cite{emmanuel_2016_pre}
In such case, the energy current depends only on the spin current
multiplied by the magnetic field. If $B_i=0$, then
$\langle J_i^E\rangle=0$ as well. A physical interpretation for
this fact is well known, see, e.g., Ref. \onlinecite{mendoza-arenas_2013}.
There, only the case in which $f=1$ is analyzed, but the result can
be extended to any $f$. The argument is that the current is given
by the superposition of two contributions, one that flows from
left to right and carries energy associated with $up$ spins
injected into site $1$ and another flowing from right to left,
carrying energy associated with $down$ spins injected into site
$N$. If no magnetic field is present in the system, these contributions
cancel each other and the energy current becomes zero.
This result does not depend on the system size, and, we stress, it does not follow for the asymmetric chain considered here. 





 \subsection{Rectification}

\begin{figure}[h!]
\centering
\includegraphics[scale=0.35]{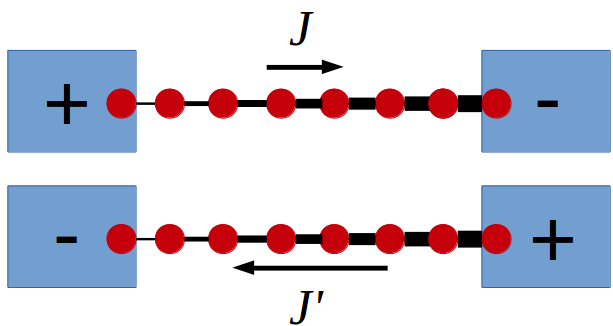}
\caption{\small{Representation of the rectification process. For
reservoirs defined by a given value of $f$ (top), a current $J(f)$
flows between them. If we invert the reservoirs by changing $f$
to $-f$ (bottom), a current $J(-f)$ appears, which has a direction
opposite to that observed in the previous case. The rectification
exists when $|J(f)| \neq |J(-f)|$.}}
\label{ret}
\end{figure}

As discussed above, if the reservoirs are not balanced,
currents usually flow along the system in a given direction. If we invert the reservoirs,
changing $f$ by $-f$ [see eqs. (\ref{lind2_1}) and (\ref{lind2_2})], the direction
of the currents also inverts, as represented in Fig.~\ref{ret}. If the chain is
symmetric, the inversion of the reservoirs does not change the absolute values of the
currents, that is, the currents are odd functions of $f$, $J(-f)=-J(f)$.
On the other hand, when the system is not symmetric, we may have $J(-f)\neq - J(f)$;
if this is the case, we say that there is rectification.

The rectification factor $R$ can be quantified as follows
\begin{equation}\label{retif}
R=100\times\frac{J(f)+J(-f)}{\left|J(f)-J(-f)\right|}.
\end{equation}
Note that for symmetric systems the rectification vanishes and that,
according to this definition, there is no bound on the values of $R$.

Recent works in the literature have focused on studying the
rectification of the spin\cite{landi_2014_pre} and the
energy\cite{emmanuel_2016_pre} currents in XXZ chains. In both cases
the authors analytically treated chains with $N=3$ sites, and performed
numerical computations for large chains.
Regarding the rectification of the spin current $(R_S)$, for asymmetry
present only in the magnetic field and small values of $f$, $B_j$ and
$\Delta_{j,j+1}=\Delta$, it has been shown that $R_S$ is proportional to
$\Delta$,\cite{landi_2014_pre} meaning that the anysotropy parameter
needs to be different from zero if rectification of the spin current
is desired. It is also known that $R_S$ does not depend on the average
external magnetic field, but it does depend on the gradient of the magnetic
field, that is, if the field is uniform, $R_S = 0$.\cite{landi_2014_pre}
Such a vanishment still follows even if there exists an asymmetry in the
exchange couplings.

Concerning the energy current, since it involves both
the XXZ term as well as the other term proportional to the magnetic field,
the presence of an asymmetry in the exchange coupling $\Delta_{i,i+1}$ or
in the magnetic field may generate rectification, whose factor we denote
by $R_E$. When $B_i=0$ but $\Delta_{i,i+1}$ varies along the chain,
the \textit{one-way street} phenomenon holds,\cite{emmanuel_2017_pre}
which means that the energy current propagates in only one direction,
even if we invert the reservoirs, that is, $J(f)=J(-f)$ (see Fig.~\ref{ret}).
The thermodynamical details and consistency of the one-way street
phenomenon are described in Ref.~\onlinecite{pereira2018heat}.

Since we are interested in obtaining rectification in both spin and
energy currents, the results mentioned above justify why we need to
consider an asymmetry in the external magnetic field. Note that, in
this case, the \textit{one-way street} phenomenon does not hold. Most
interestingly, because of the two
terms in Eq.~(\ref{cor-energ}), we can properly choose the asymmetry
in $\Delta_{i,i+1}$ (which defines the contribution $\langle J^{XXZ}
\rangle$ to $\langle J^E \rangle$) and that of $B_j$ (to which
$\langle J^{B} \rangle$ is proportional) in such a way that the energy
current has opposite direction with respect to the spin current, as
we explore in the next section.

\section{Results}

We present now our results for the spin and energy currents flowing
through the system described by eq. (\ref{xxz}) to (\ref{lind2_2}).
We explore the possibility of manipulating and controlling the
currents and rectifications (1) by properly building the systems, which means
by properly choosing the exchange couplings $\Delta_{j,j+1}$ defining the
chain, or (2) by performing external interventions, through the application
of an inhomogeneous external magnetic field $B_{j}$. We also analyze the
behavior of rectification when the system degree of asymmetry changes or
the chain is chosen to be in different ``phases'', i.e., part of the chain has $-1<\Delta<1$ and part has
$|\Delta|>1$. 



The behavior of the currents is very sensitive to the values of the involved
parameters $\varepsilon$, $f$, $\Delta_{j,j+1}$, and $B_{j}$. Hence, as a first study,
we concentrate on specific situations in which we observe reversal of the energy
current and its rectification through the application of an external magnetic field.
In particular, we take fixed values for some parameters, $\varepsilon = 1$ and
$|f| \sim 0.5$, and focus on the properties due to changes and asymmetries in
$\Delta_{j,j+1}$ and $B_{j}$.

The asymmetries in $\Delta_{j,j+1}$ and $B_j$ are quantified by $\delta$ and $\zeta$,
respectively, as follows
\begin{equation}\label{linear_eq}
  \Delta_{i,i+1} \in \{\Delta-\delta, \cdots, \Delta, \cdots, \Delta+\delta\},
\end{equation}
\begin{equation}\label{incremento_eq}
 B_i \in \{\cdots,h-2\zeta,h-\zeta,h,h+\zeta, h+2\zeta, \cdots\}.
\end{equation}

\noindent
In the first type of asymmetry, the parameters at the chain
edges are given by two fixed values, $\Delta-\delta$
and $\Delta+\delta$, and the parameters along the chain
are linearly distributed in between these two edge values. 
In the second type, for $B$ above, an
incremental equal to $\zeta$ is subtracted (added) to the central
parameter, $h$, as one moves along the chain to the left (right), beginning at
the central site. Note that, in the case of the asymmetry occurring 
in the inter-site interaction $\Delta$ instead of in the external field $B$, 
each term involves two sites. Since in the first type of asymmetry the edge 
values are fixed once $\delta$ is defined, in contrast to what happens in the
second type, we name the former situation as a delimited asymmetry, while the 
last one is denominated as incremental asymmetry.

Here, we obtain the NESS for chains up to $N=7$ through exact diagonalization.
For chains with $N>7$, $\rho_{\infty}$ is calculated by means of the time evolution,
precisely, by using time dependent density matrix renormalization group
(t-DMRG) \cite{schollwock2011density, verstraete2008matrix} techniques, where the largest bond-dimension used was 120. We have
checked that for $N\leq 7$ exact diagonalization and t-DMRG computations agree with
each other. The criterion used to assure that the NESS was reached is the observation
of homogeneity in the spin and energy currents along the chain. We consider that the currents are homogeneous when their site to site variations are of the order of $10^{-5}$, in the case of the spin current, and of the order of $10^{-4}$, in the case of the energy current, which is reached for chains of up to 9 sites.



We present our results and discuss them in more detail below.


\subsection{Behavior as a function of the anisotropy parameter} \label{depDelta}

To clearly depict the scenario,  we first take a small system of $N=3$ sites, set the magnetic field to zero and analyze the currents as a
function of the exchange coupling $\Delta$ (average for the values of $\Delta_{i,i+1}$ in the chain).
Some results in this initial analysis are numerical confirmations of
statements previously established, mainly by symmetry arguments in the Lindblad master equation related to the
time evolution of the XXZ chain. \cite{emmanuel_2017_pre, landi_2014_pre} These statements essentially concern existence or vanishment of rectification in the system.
In a following step, we turn on the magnetic field and give the first examples of changes in the scenario due to external interference.
 We describe these and other findings below.

\begin{figure}[ht]
\centering
\includegraphics[scale=0.5]{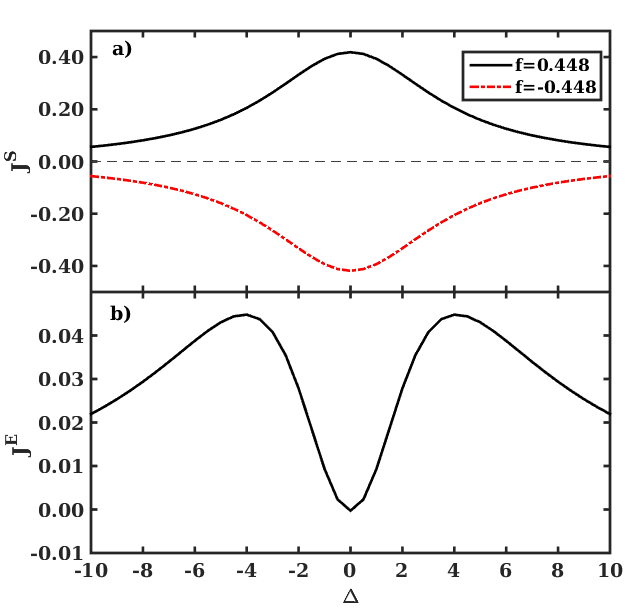}
\caption{\small{(a) Spin and (b) energy currents as a function of $\Delta$ (average
    value of $\Delta_{i,i+1}$)
    for a chain of $N=3$ sites. We consider the asymmetry in $\Delta_{i,i+1}$, set by $\delta=0.15$. There is no external
    magnetic field (that is, $h=0.0$ and $\zeta=0.0$) and the baths
    are characterized by $|f|=0.448$.}}\label{mani01}
\end{figure}

 In Fig.~\ref{mani01} we present the
results for the spin and the energy currents in the case of $|f|=0.448$ [parameter that
defines the baths, see eqs. (\ref{lind2_1}) and (\ref{lind2_2})] and
$\delta=0.15$, parameter which sets the asymmetry in $\Delta_{i,i+1}$, see eq. (\ref{linear_eq}). 

In panel (a) of Fig. \ref{mani01}, we see the spin current: the black continuous
line corresponds to $f=0.448$, while the red dashed
curve is the result for $-f$, that is, it is obtained when we invert the
reservoirs. We observe that $\langle J^S(f) \rangle= -\langle J^S(-f)
\rangle$, meaning that there is no
rectification. This is expected and can be rigorously proven whenever there is no asymmetry in the magnetic field, \cite{landi_2014_pre} even if there is some asymmetry in $\Delta_{i,i+1}$. \cite{emmanuel_2017_pre}
In panel (b), we have the energy current.
Here we observe the \textit{one-way street} phenomenon,\cite{emmanuel_2017_pre}
that is, the two curves, one for $f=0.448$ and another for $f=-0.448$, coincide.
It means that the current flows in the same direction irrespective of the
reservoir configuration, precisely, the energy current is an even function of $f$.
This can be proven to hold \cite{emmanuel_2017_pre} whenever the external magnetic field vanishes; however, if it does not vanish,  one term proportional to the magnetic field appears and the two currents become different.



\begin{figure}[h!]
 \centering
 \includegraphics[scale=0.5]{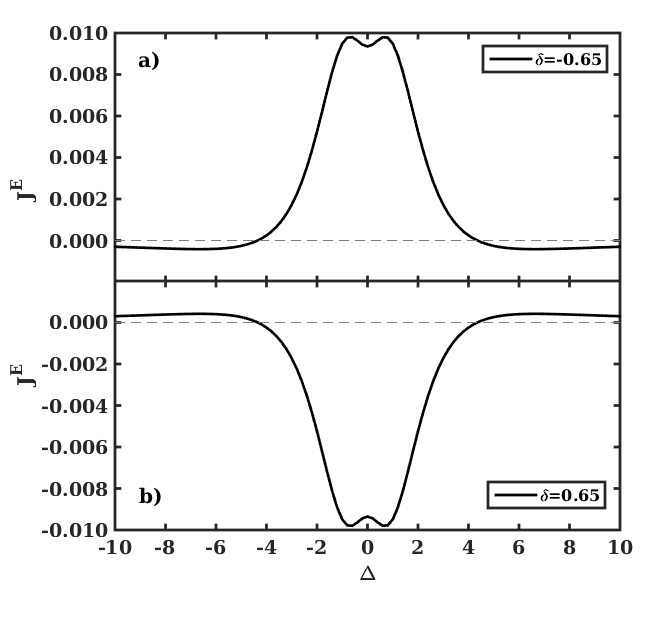}
 \caption{\small{Energy current as a function of $\Delta$ for (a)
     $\delta=0.65$ and (b) $\delta=-0.65$. Other parameters as in
 Fig.~\ref{mani01}.}}\label{d065}
 \end{figure}

In Fig.~\ref{mani01} we can also observe that the energy current, in the
absence of an external field, is an even function of $\Delta$, that is,
$\langle J^E(\Delta) \rangle= \langle J^E(-\Delta) \rangle$.
However, the detailed dependence of the current on
 $\Delta$ and $\delta$ is not trivial. For example, we
cannot say that the direction of the current is given by the gradient of
$\Delta$: in other words, $\delta>0$ does not imply that the current
is positive. Indeed, the results can vary considerably with $\delta$, as exemplified by the data in Fig.~\ref{d065}, for
$\delta = 0.65$ [panel
(a)] and $\delta = -0.65$ [panel (b)]. Note that the current
sign depends on the anisotropy $\delta$. Moreover, the values of
$\Delta$ for which the current is maximum change in comparison with
the previous case (Fig.~\ref{mani01}). Finally, we can see another symmetry:  $\langle J^E(\delta) \rangle = -\langle J^E
(-\delta) \rangle$.

Let us now turn on the external magnetic field
applied to the system. Now the energy current may depend on the sign of $f$. This happens because the term
of the energy current that is proportional to $B_i$ ($\langle J_i^B \rangle$
given in eq.~(\ref{campo})) depends also on the spin current, and, thus, it can
give a positive or negative contribution to $\langle J^E \rangle$, depending on the sign of $f$. It can lead to the emergence of energy rectification. However, the rectification in the spin current does not
appear if we simply turn on $B_i$: its occurrence is dependent on the existence of asymmetry in $B_i$, here, defined  by $\zeta$.

\begin{figure}[ht]
 \centering
 \includegraphics[scale=0.5]{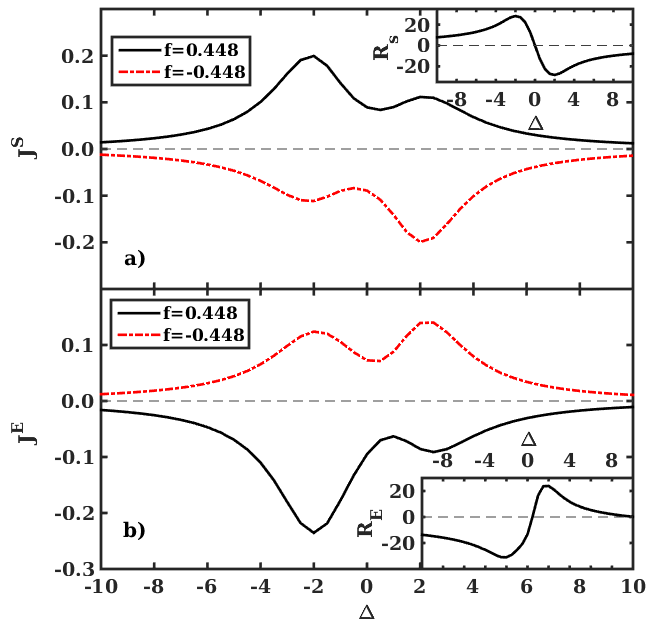}
 \caption{\small{(a) Spin and (b) energy currents as a function of $\Delta$
    for a chain of $N=3$ sites. We consider the asymmetry in $\Delta$, set by $\delta=0.15$,
    and the asymmetry in $B_i$, defined by $h=-0.94$ (average value of
    $B_i$) and $\zeta=-2.0$.
    The baths are characterized by $|f|=0.448$. The
    insets show the rectifications calculated from the results in the main
    panels.}}\label{mani02}
 \end{figure}

In Fig.~\ref{mani02}, we show the results for the
spin and energy currents as a function of $\Delta$, and in the presence of
asymmetries in $\Delta_{i,i+1}$ and in $B_i$. The curves for the currents present expressive modifications as we turn on an asymmetric magnetic field  $B_i$
(compare Fig.~\ref{mani01} and~\ref{mani02}). In panel (a) we see
that $\langle J_S(f) \rangle$ is not equal to
$-\langle J^S(-f) \rangle$ anymore and thus there is
rectification of the spin current. The differences are more significant
in panel (b), where the coincidence between the curves for $\langle J^E (f) \rangle$ and
$\langle J^E (-f) \rangle$ disappears. In both panels, there are new values for $\Delta$  maximizing the currents:
for $B_i=0$  the maximums are in  $\Delta= \pm 4$, while, in the presence of $B_i$, the maximums
follow for $\Delta=2$, when the energy current propagates
to the right, and $\Delta=-2$, when it propagates to the left.

It is interesting to
note that the spin and the energy currents propagate in opposite directions for
a given configuration of the reservoirs: for $f$ positive (negative),
the spin current is positive (negative), while the energy current is
negative (positive). Since the spin current does not depend on the average
magnetic field $h$, it means that we can control, and even invert, the direction of
the energy current via the choice for the values of the external magnetic field. This is a clear example of procedure allowing
the manipulation of the currents through external mechanisms.

 As can be  clearly noticed in Fig.~\ref{mani02}, the results are not symmetric with
 respect to the positive/negative $y$-axis, in opposition to the results in Fig.~\ref{mani01}. It means
  that we have rectification for both currents. In the insets of Fig.~\ref{mani02}, we show the
 rectification factors calculated from the currents.
 For example, if we fix $\Delta=4$,  we observe that the
spin current flows more easily from the right to the left side (negative $R_S$), while the
energy current runs more easily in the opposite direction (positive $R_E$).



\begin{figure}[ht]
 \centering
 \includegraphics[scale=0.5]{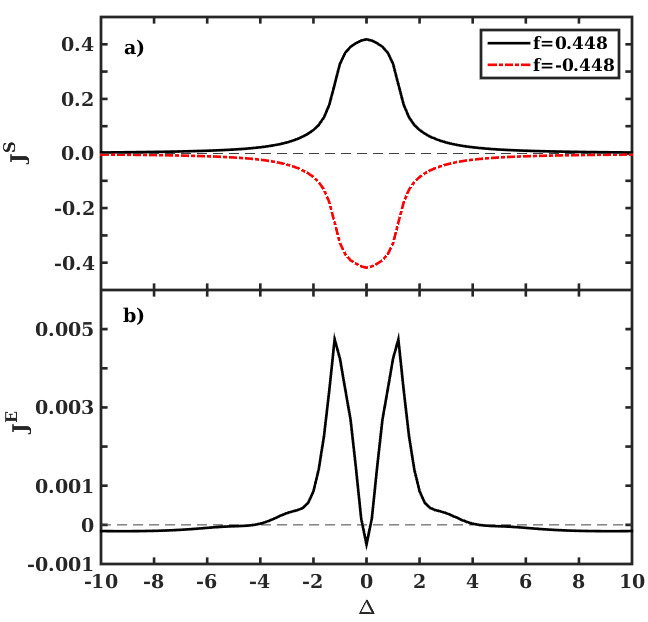}
 \caption{\small{(a) Spin and (b) energy currents as a function of $\Delta$ for a chain of $N=7$ sites.
 We consider the asymmetry in $\Delta_{i,i+1}$, set by $\delta=0.15$}. There is not an external
    magnetic field applied (that is, $h=0.0$ and $\zeta=0.0$) and the baths
    are characterized by $|f|=0.448$.}\label{N7_h0_lin}
 \end{figure}

\begin{figure}[ht]
 \centering
 \includegraphics[scale=0.5]{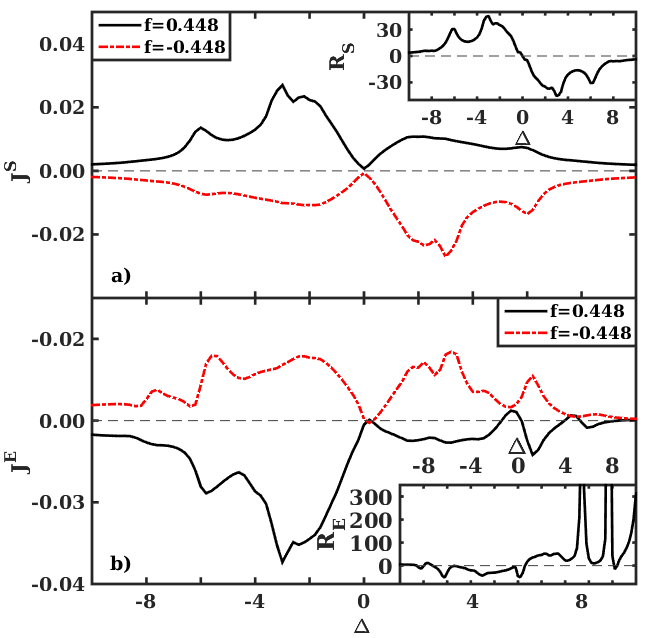}
 \caption{\small{(a) Spin and (b) energy currents as a function of $\Delta$
    for a chain of $N=7$ sites. We consider the asymmetry in $\Delta_{i,i+1}$, set by $\delta=0.15$,
    and the asymmetry in $B_i$, defined by $h=-0.94$ (average value of
    $B_i$) and $\zeta=-2.0$.
    The baths are characterized by $|f|=0.448$. The
    insets show the rectifications calculated from the results in the main
    panels.}}\label{N7_h-094_lin}
 \end{figure}

Now we extend the previous investigation to larger chains, mainly $N=7$, and describe also some new phenomena.
Following the previous script, we first compute
the spin and the energy currents in a chain with zero external magnetic field (Fig. \ref{N7_h0_lin}) and, after that, we turn on the magnetic field along the chain (Fig. \ref{N7_h-094_lin}) and compare the results.
As before, we plot the currents as a function of $\Delta$. It is important to note that, if we fix the asymmetric parameter $\delta$ at the edges of the chains,
 as we increase the system size, the asymmetry in $\Delta_{i,i+1}$ becomes smoother. Therefore, a chain with $N=7$ is less asymmetric than a chain with $N=3$.


 Panels (a) and (b) in Fig. \ref{N7_h0_lin} present the spin and energy currents, respectively, for a chain with no magnetic field applied. Again, there is no rectification for the spin, and
 the \textit{one-way street} phenomenon \cite{emmanuel_2017_pre} remains for the energy current, as expect, since these results do not depend on the chain size. Note that, in a comparison with the case $N=3$, the behavior of the energy current changes, and the maxima go to $\Delta \simeq \pm 2$ instead of $\Delta= \pm 4$.

  In Fig. \ref{N7_h-094_lin} we turn on the magnetic field along the chain. The asymmetry in the field is given by Eq. \ref{incremento_eq}, where we use $h=-0.94$ and $\zeta=-2$. Now, the curves significantly change and
  we observe the occurrence of rectification for both spin and energy currents, see the insets. It is interesting to note a huge rectification in the energy current for some values of $\Delta$, e.g., $\Delta\approx 5$ and $\Delta\approx 8$.
For a given reservoir configuration and for most of the $\Delta$ values, the energy and the spin currents propagate in opposite directions, see Fig.~\ref{N7_h-094_lin}.  We obtain this effect
  by turning on the external magnetic field with properly chosen value and sign. The direction of the spin current is defined by the reservoirs configuration,
 however the direction of the energy current, see eq. (\ref{cor-energ}), involves the magnetic field, and we can adjust it
 to manipulate the energy flow.

\begin{figure}[h!]
 \centering
 \includegraphics[scale=0.52]{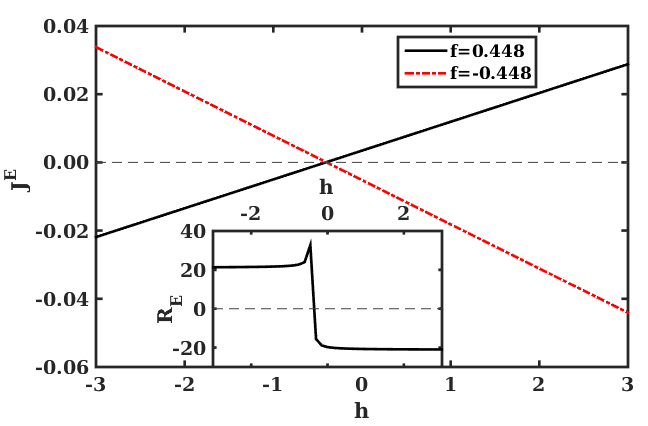}
 \caption{\small{Energy current as a function of $h$ (average value of $B_i$)
    for a chain of $N=7$ sites. We $\Delta=4$ and the asymmetry in $\Delta_{i,i+1}$, set by $\delta=0.15$,
    and the asymmetry in $B_i$, defined by $\zeta=-2.0$.
    The baths are characterized by $|f|=0.448$. The
    inset shows the rectification calculated from the results in the main
    panel.}}\label{ecvsh}
 \end{figure}

 In Fig.~\ref{ecvsh},  for a fixed value of $\Delta$, which we take as $\Delta=4$,  we analyze  the
behavior of $\langle J^E \rangle$ as a function of $h$, the average value of $B_i$. As we can see, it is possible to change the propagation direction of the energy current by varying $h$.
In contrast, the spin current does not depend on $h$, and its direction is given by the reservoir configuration. Thus, we have seen that, 
taking a system with a properly chosen and constant structure, we can use the
external magnetic field to obtain, for example, the spin and the energy currents flowing in opposite directions.

\subsection{Changing the degree of asymmetry}

We now analyze situations in which the system degree of asymmetry changes.
This is the case when we increase the chain size $N$, keeping the values of
$\Delta$ and $\delta$ fixed, given the delimited type of asymmetry we
consider for the anisotropy parameter [see eq.(\ref{linear_eq})]; as
previously mentioned the
system asymmetry indeed becomes smoother as $N$ increases. If on the other hand we
keep $\Delta$ and $N$ fixed, we can obtain a more asymmetric system by increasing
$\delta$. Results for these two cases are discussed in the present subsection.

\begin{figure}[h!]
 \centering
 \includegraphics[scale=0.52]{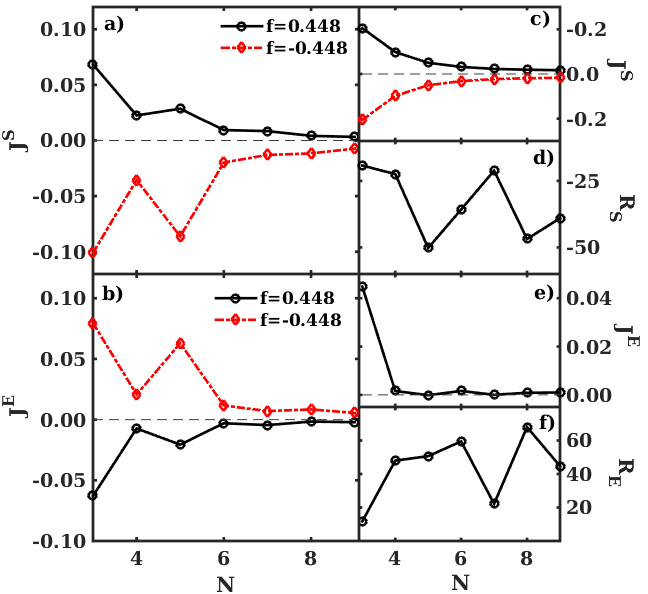} \\
 \caption{\small{(a) Spin and (b) energy currents as a function of $N$. The parameters used
     were $\Delta=4$, $\delta=0.15$, $h=-0.94$, $\zeta=-2.0$, and
     $|f|=0.448$. (c) Spin and (e) energy currents as function of $N$ in the absence of external magnetic field. (d) and (f) show the rectifications calculated from the results in panels (a) and (b), respectively.}}\label{linear}
 \end{figure}

 In Fig. \ref{linear} we plot the spin current, panel (a), and the energy current, panel (b), as a function of $N$ for $\Delta=4$
in the presence of an asymmetric external magnetic field [the results in the absence of the field are shown in
panels (c) and (e) for comparison.] For the chosen selection of parameters, i.e., in the analyzed regime,
 both the spin and the energy currents decrease in magnitude as the chain size $N$ increases. However, the corresponding rectification
factors (shown in panels (d) and (f)) oscillate but do not decay. We note that, due to the denominator in the definition of
$R$ in eq.~\eqref{retif}, this quantity can be highly sensitive to small variations in the model parameters, such as the system size,
especially when the currents in both directions are small.

 We also observe, in all these chains, the appearance of previously  described phenomena, stimulated by arrangements between the inner system structures and the manipulated external magnetic field. Namely,
   we are able to invert the  propagation direction of
the energy current in comparison with that of the spin flow. Consequently, we can obtain inverted rectification signs for the energy and spin currents.
It is important to emphasize that such phenomena only take place when there is an
asymmetry in the external magnetic field and asymmetry in the inner structure of the system.

\begin{figure}[h!]
 \centering
 \includegraphics[scale=0.5]{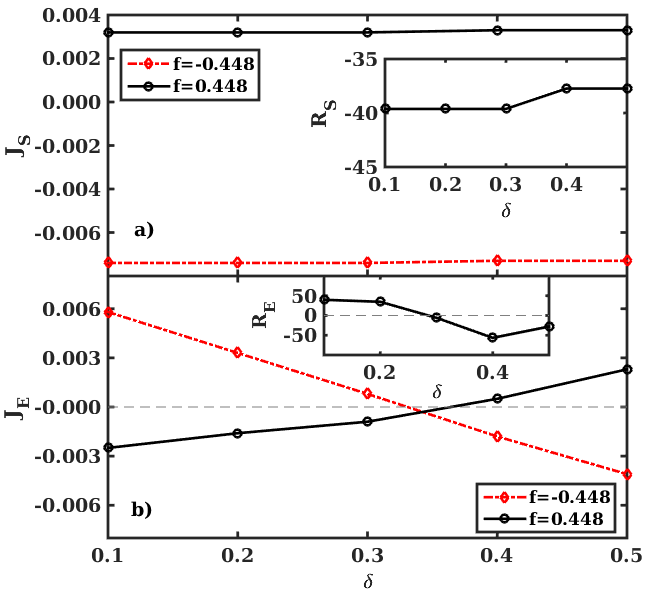}
  \caption{\small{(a) Spin and (b) energy currents as a function of $\delta$
     for the delimited asymmetry in $\Delta_{i,i+1}$ and a chain with $N=9$ sites.
     The other parameters used were $\Delta=4$, $h=-0.94$, $\zeta=-2.0$, and
     $|f|=0.448$. The insets show the rectifications calculated from
     the results in the main panels.}}\label{delta}
 \end{figure}

We find a further, surprising effect in the energy current due to the asymmetry in the inner parameter $\Delta_{i,i+1}$. For the regime of parameters that we have used, and for a chain with $N=9$ sites, as the asymmetry
$\delta$ increases,  the energy current inverts its direction while the spin current remains essentially constant, see Fig.~\ref{delta}.
For concreteness, for the given set of parameters $\Delta, h, \zeta$, with fixed reservoirs, the energy current propagates in one direction for $\delta=0.15$ and flows in the opposite direction for $\delta=0.5$. This inversion of the current direction implies in a change of the corresponding rectification sign, see the inset of panel (b) in Fig \ref{delta}. We observe this effect in another system size, for example, in chains with $N=7$ sites.

\subsection{Chains in different ``phases''}\label{phase}

It seems to be propitious the investigation of the system behavior for $\Delta$ corresponding to a chain in the critical phase ($|\Delta| < 1$),
 for a comparison with $\Delta$ in the anti-ferromagnetic phase ($\Delta > 1$), as considered in some of the previous computations. We use the nomination phase here, although it is appropriate only in the thermodynamic limit.
For a clear  comparison, we depict the curves for two values: $\Delta=0.5$ and $\Delta=1.5$. Moreover, we take $|f|=0.5$, $\delta=0.25$, $h=0.1$ (to avoid the \textit{one-way street} phenomenon), and $\zeta=0$.

\begin{figure}
   \centering
 \includegraphics[scale=0.5]{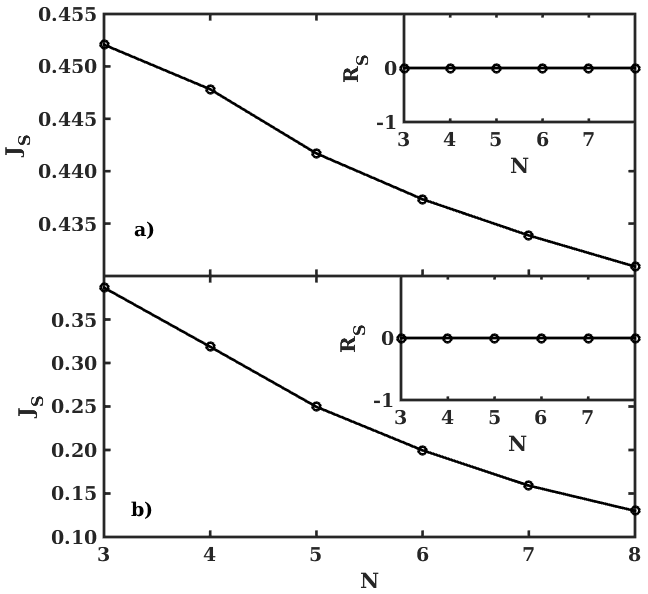}
 \caption{\small{Spin current as a function of the chain size $N$ for (a)
$\Delta = 0.5$ and (b) $\Delta = 1.5$. We choose $\delta = 0.25$, $h = 0.1$,
$\zeta = 0$ (no asymmetry in the external magnetic field) and $f = 0.5$.
The insets emphasize that in both cases, if we invert the reservoirs
(not shown), the rectification factors vanish, since the currents do not
change.}}\label{spin}
 \end{figure}

In Fig.~\ref{spin}, we  present the spin current as a function of the chain
size $N$ for $\Delta=0.5$, panel (a), and $\Delta=1.5$, panel (b). The current decreases with the
system size in both cases, but the decay rate is larger for $\Delta=1.5$
than for $\Delta=0.5$. As well known in the literature, for homogeneous systems,  the spin current is ballistic in the critical phase: to be specific,
 no dependence with the chain size has been observed
in the case corresponding to $\Delta = 0.5$,  for a system in the
absence of asymmetry and for chains larger than $20$.
\cite{prosen2009matrix} On the other hand, for $\Delta>1$,
the spin transport is diffusive, in other words, it decays with the chain size.~\cite{prosen2011exact}

\begin{figure}[h!]
   \centering
 \includegraphics[scale=0.5]{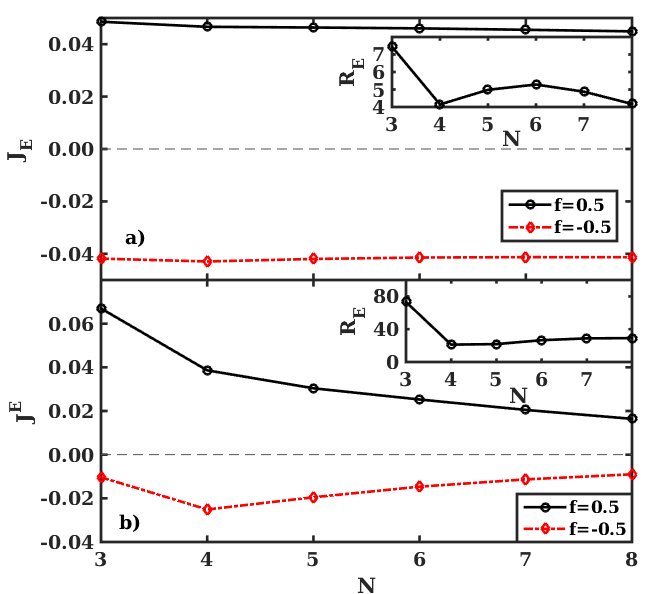}
 \caption{\small{Energy current as a function of the chain size $N$ for
(a) $\Delta = 0.5$ and (b) $\Delta = 1.5$, when the reservoirs are defined by
$f = 0.5$ and when they are inverted $(f = -0.5)$. Other parameters
as in Fig. \ref{spin}. The insets show the corresponding rectification factors.}}\label{energia}
 \end{figure}

In Fig.~\ref{energia}, we depict the dependence of the energy currents with the chain size,
for the same set of parameters as those considered in Fig.~\ref{spin}. We see that $\langle J^E \rangle$ slightly
changes in the case of $\Delta=0.5$, in contrast with the decay
observed for $\Delta=1.5$.

 The rectification factors are shown in the insets of both Fig.~\ref{spin}
and Fig.~\ref{energia}. The rectifications
vanish for the spin currents, as expected (homogeneous magnetic field), but not for the energy flows.
An interesting point is that $R^E$ is much larger for $\Delta= 1.5$ than for $\Delta=0.5$.


\section{Conclusions}

In this paper we consider a graded spin-1/2 chain, described by the
Heisenberg XXZ model and boundary driven by magnetic baths. The
asymmetry present in the system can be either intrinsic to it,
such as in the exchange coupling in the $Z-$direction, or external,
as given by the application of a magnetic field that changes from site
to site. We analyze the spin and energy currents, as well as the
respective rectifications. According to our results, it is possible
to control and manipulate the energy current through the application
of an external magnetic field. One has, though, to carefully choose
the parameters that define the system, since the behavior depends
strongly on them. If this is done, it is possible that the spin
current propagates in one direction, while the energy current propagates
in the opposite direction, if the reservoir is kept fixed. When the
reservoirs are inverted, both currents change direction. In this situation,
if the rectification of the spin current is positive that of the
energy current is negative.
Another interesting we have observed is that of an inversion of the energy current
direction as the asymmetry present in the exchange
parameter increases. It will be the subject of future work to apply appropriate numerical techniques to extend the results to larger chains.

It is pertinent to recall that it is possible to engineer XXZ Hamiltonian spin systems
with different values for the parameters $\Delta$ and $\alpha$,\cite{endres2016atom,barredo2016atom}
and that related XXZ models are involved in recent experiments with Rydberg atoms in optical
traps.\cite{duan2003controlling,whitlock2017simulating,PhysRevX.8.011032} In the present work, we propose a many-body spin device, based in
graded XXZ chains, in which, besides the recalled previous effects, it is possible to induce the reversal of energy rectification by simply introducing properly chosen
external magnetic fields, precisely, it is possible to invert the direction of the bigger energy flow by means of an external magnetic field. 

We acknowledge Eduardo Mascarenhas and Gabriel Landi for providing
us their MPS (without asymmetries) and exact diagonalization codes,
respectively. We thank CNPq, CAPES, and FAPEMIG for financial support;
support from the INCT on Quantum Information/CNPq is also gratefully
acknowledged.


%
%

\bibliography{refer}

\end{document}